# Double-negative acoustic metamaterial based on hollow steel tube meta-atom

Huaijun Chen, Hongcheng Zeng, Changlin Ding, Chunrong Luo and Xiaopeng Zhao

Smart Materials Laboratory, Department of Applied Physics,

Northwestern Polytechnical University, Xi' an 710129, People's Republic of China

E-mail: xpzhao@nwpu.edu.cn

**Abstract**

We presented an acoustic 'meta-atom' model of hollow steel tube (HST). The simulated and experimental results demonstrated that the resonant frequency is closely related to the length of the HST. Based on the HST model, we fabricated a two-dimensional (2D) acoustic metamaterial (AM) with negative effective mass density, which put up the transmission dip and accompanied inverse phase in experiment. By coupling the HST with split hollow sphere (SHS), another kind of 'meta-atom' with negative effective modulus in the layered sponge matrix, a three-dimensional (3D) AM was fabricated with simultaneously negative modulus and negative mass density. From the experiment, it is shown that the transmission peak similar to the electromagnetic metamaterials exhibited in the double-negative region of the AM. We also demonstrated that this kind of double-negative AM can faithfully distinguish the acoustic sub-wavelength details ( $\lambda/7$ ) at the resonance frequency of 1630 Hz.

*September 27, 2012*

# 1. Introduction

As a kind of artificial material, left-handed metamaterials (LHMs)[1-3] have realized some unique properties not found in nature, such as sub-wavelength imaging, negative refraction and invisibility cloaking.[4-6] Considering the similarity between electromagnetic waves and acoustic waves, the analytical methods and theories for LHMs are suitable for acoustic metamaterial (AM) to successfully realize the similar unique properties.[7-10] Guided by the concept of local area resonance, some different artificial acoustic 'atoms' are engineered and used to fabricate AM to realize negative acoustic parameters. Researchers achieved a three-dimensional (3D) negative mass density AM using the model involving a solid core coated with an elastic soft material[11] and then developed a two-dimensional (2D) membrane-type AM with negative density.[12] Moreover, a series of AM with the sub-wavelength helmholtz resonators (HR) were designed to achieve negative modulus and even double-negative parameters based on transmission line theory.[13,14] In theoretically, AM with simultaneously negative mass density and negative modulus can be fabricated by coupling the two kinds of acoustic atoms,[15,16] and in fact it is an important way to realize double-negative parameters. Up to now, the one-dimensional double-negative AM has been experimentally achieved.[17] However, the difficult coupling between two kinds of AM structures and the mismatch for their resonant frequencies led to the current situation that double-negative AM remains in the simulation stage and can hardly be experimentally fabricated.

Such difficulties are largely solved in this paper. In our early study, we proposed a split hollow sphere (SHS) model, a kind of artificial acoustic 'meta-atom' which can experimentally achieve negative effective modulus at the audible sound frequency by periodically arraying SHS in the sponge matrix.[18,19] The significant advantage of this SHS model is that it can be easily coupled with other negative effective mass density AM to realize double negative, and the negative frequency can be adjusted by varying both the ball and hole diameters on the SHS. In this paper, we propose a hollow steel tube (HST) model, another kind of artificial acoustic 'meta-atom', which 2D structure can experimentally realize negative mass density and the frequency of negative mass density can be adjusted by changing the HST length. This HST model has the simple structure and can easily be coupled with other acoustic 'atoms' to construct 2D or 3D AM. The presented characteristics of SHS and HST allow the experimental realization of double-negative AM via coupling with each other.

## 2. The proposition and realization of negative effective mass density

### A. Model construction and theoretical analysis

Based on the transmission line theory, a single HST can be equivalent to *L-C* resonant circuit. Fig. 1 shows the structural scheme of a single HST and its equivalent circuit. The HST has a diameter $d$, wall thickness $t$, and length $l$. In the equivalent circuit, the capacitance $C$ and the inductance $L$ are proportional to the acoustic capacitance $C_a$ and to the acoustic mass $M$, respectively. The voltage $V$ above the *L-C* resonant circuit corresponds to the acoustic pressure, and the current $I$ corresponds to the sound velocity.

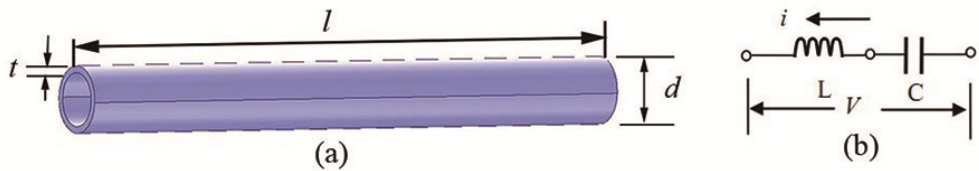

Figure 1: (a) Schematic and (b) equivalent *L-C* circuit of the single HST.

According to Ohm's law, the mathematical relation between the current $I$ and the voltage $V$ can be obtained, as shown in Eqs. (1).

$$V = L\frac{dI}{dt} + \frac{1}{C}\int I dt, \tag{1}$$

Considering that the HST is a homogeneous tube and $I$ in the $L$-$C$ circuit is time-harmonic, Eqs. (1) can be rewritten as

$$\nabla v = (i\omega L + \frac{1}{i\omega C})\vec{j} = i\omega L_1 \vec{j}, \tag{2}$$

where $\vec{j}$ is the current density, $v$ is the voltage distribution, $L_1$ is the effective inductance of the circuit. From Eqs. (2), we can derive

$$L_1 = L(1 - \frac{\omega^2_1}{\omega^2}), \tag{3}$$

where $\omega^2_1 = \frac{1}{LC}$.

From the corresponding physical relation between the $L$-$C$ circuit and the HST model, we can determine that the effective inductance $L_1$ of the circuit corresponds to the effective mass density. Thus, we obtain

$$\rho_{eff} = \rho_0(1 - \frac{\omega_1^2}{\omega^2}), \tag{4}$$

Where $\rho_{eff}$ is the effective mass density of the HST, $\rho_0$ is the air mass density, $\omega_1$ is the air resonance frequency in the HST, and $\omega$ is the acoustic frequency. By assuming that the cavity of the HST can be simultaneously regarded as acoustic inductor and acoustic capacitor, the resonance frequency $\omega_1$ can be obtained as follows

$$L = \frac{\rho l}{\pi(\frac{d}{2} - t)^2},$$

$$C = \frac{\pi(\frac{d}{2}-t)^2 l}{\rho c^2},$$

$$\omega_1^2 = \frac{1}{LC} = \frac{c^2}{l^2}, \tag{5}$$

Eqs. (4) and (5) show that the resonance frequency and negative mass density are closely related to the length of the HST.

## B. The experimental realization of the negative effective mass density

To verify the effective mass density of the AM with single HST, the amplitudes and phase curves of transmission and reflection are experimentally measured. The HST possesses certain parameter values ($d$=10 mm, $t$=0.4 mm, and $l$=95 mm) and is pasted on a cylindrical sponge matrix with a thickness of 30 mm and a radius of 50 mm. This cylindrical sponge matrix has been proven suitable as a sound matrix.[18,19]

The measuring device is the impedance tube system in which the frequency of the acoustic wave is scanned in the range of 400 Hz to 1930 Hz at 2 Hz intervals. The test sample is placed at the middle of the impedance tube system, in which the transmission amplitude and the phase can be acquired using four microphones distributed at fixed positions in the tube. Before testing the reflection, the absorbing material is used at the end of the impedance tube system to absorb transmission sound. By using the two-microphone method, the refection amplitude and phase curves are obtained.

Fig. 2 shows the experimental results and the numerical simulations of the transmission amplitude and phase curve using the finite element method (FEM). The numerical simulations match well with the experiment results. A deep transmission dip near 1620 Hz exhibits in the measured transmission amplitude curve of the model in which the transmission ratio is -5.45 dB. The transmission phase curve of the model has a phase fluctuation at the dip frequency, which indicates that the low transmission ratio is caused by air resonance in the HST. The transmission dip and inverse phase that simultaneously appeared near 1620 Hz are the symbols of the effective negative density.

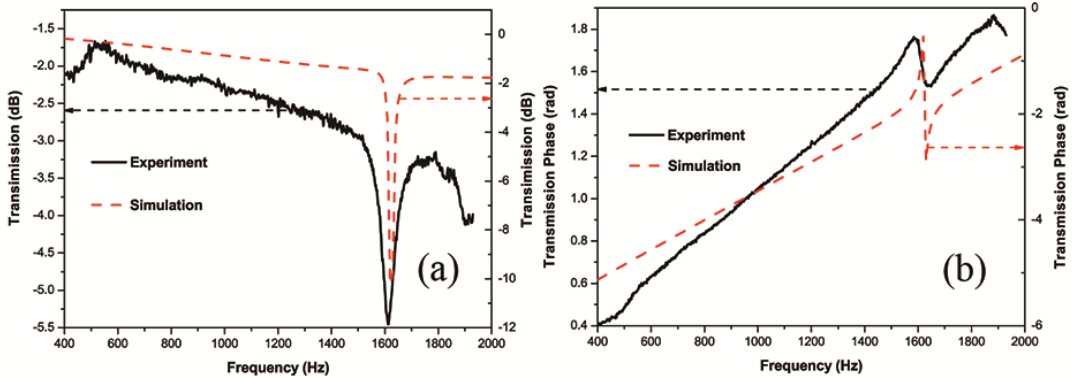

Figure 2: (a) Transmitted amplitude and (b) transmitted phase for the AM with single HST from the simulation and experiment.

Fig. 3 shows the experimental results of the reflection amplitudes and phase curves of the single HST AM. Owing to the fact that the lattice is much smaller than the acoustic wavelength, the HST AM can be considered as homogeneous medium. Based on the method for retrieving effective parameter from homogeneous medium, [20] we can calculate the effective density of HST AM. From Ref. 20, the acoustic refractive index $n$ and impedance $Z_{eff}$ can be obtained as follows:

$$n = \frac{\pm \cos^{-1}\{\frac{1}{2T}[1-(R^2-T^2)]\}}{kd} + \frac{2\pi m}{kd}, \tag{6}$$

$$Z_{eff} = \pm\sqrt{\frac{(1+R)^2 - T^2}{(1-R)^2 - T^2}}, \tag{7}$$

Where $R$ and $T$ are the coefficient of reflection and transmission, $k$ is the acoustic wave vector, $d$ is the thickness of the HST AM, and $m$ is the branch number of $\cos^{-1}$ function.

It can be seen that from Eqs. (6) and (7), the effective mass density can be obtained,

$$\rho_{eff} = nZ_{eff}\rho_0, \tag{8}$$

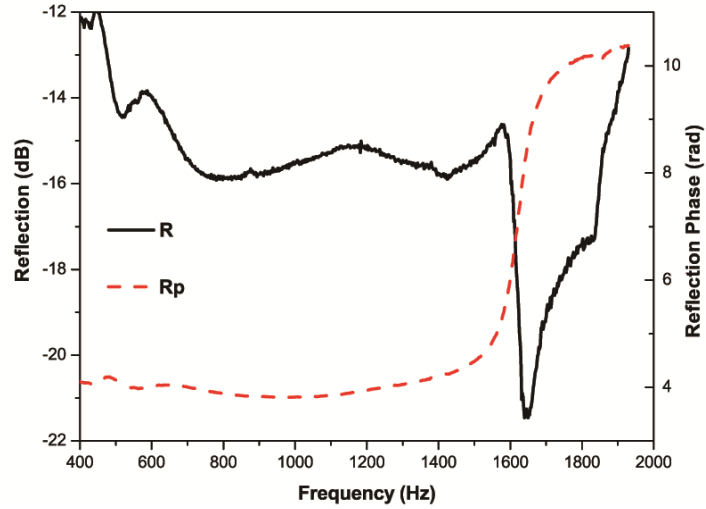

Figure 3: The reflective ratio and phase of the single HST AM.

Fig. 4 shows the effective mass density of the single HST AM. The minimum value of the real part of the effective mass density appears near the resonance frequency, which demonstrates that the minimum value is the result of air resonance in the HST. However, the air resonance in the single HST is too weak to achieve negative mass density.

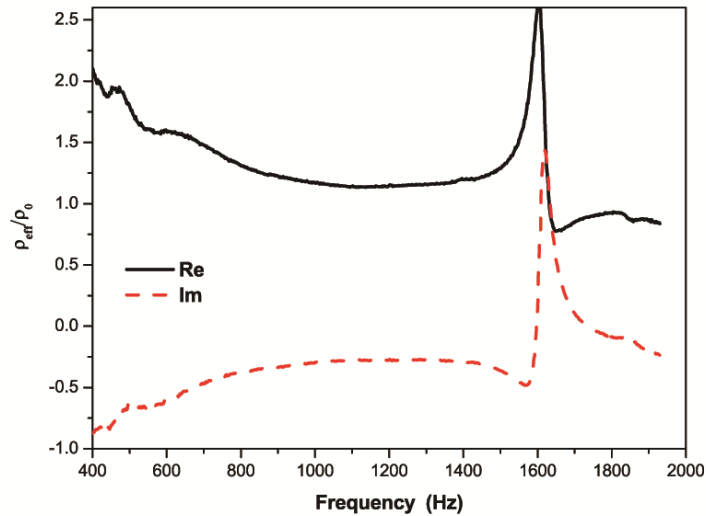

Figure 4: The effective mass density of the AM with single HST.

To strengthen the intensity of air resonance to acquire negative effective mass density, a type of 2D AM is fabricated by pasting double layers of parallel HSTs on the two surfaces of a cylindrical sponge matrix. Each layer contains nine HSTs, with the middle HST being the longest; the other eight are arranged symmetrically and are of the same length. From middle to end, the lengths of the HSTs are 96, 91, 82, 66, and 36 mm, and the thickness and the radius of the sponge matrix are 40 and 50 mm, respectively, as shown in Fig. 5(a). All HSTs possess the same diameters and thicknesses as shown in Fig. 1(a). As deduced from the above analysis, each layer can correspond to a series of resonant circuits with different parameters of $L$ and $C$ as shown in Fig. 5(b).

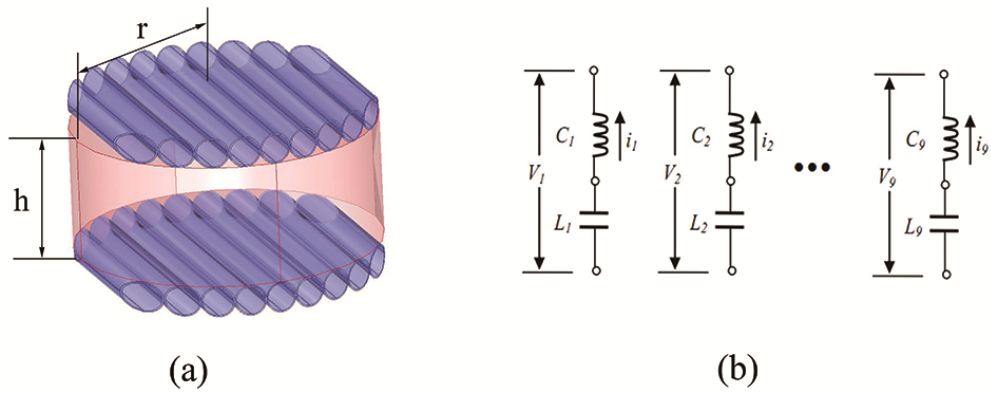

Figure 5: (a) Schematic and (b) equivalent $L$-$C$ circuit of the double layers HSTs.

The 2D AM is tested in the impedance tube system. As the frequency increases from 400 Hz to 1930 Hz, changes are observed in the trend of amplitude and in the phase curves of transmission and reflection, as shown in Fig. 6. The experimental results clearly agree with the numerical simulations. The transmission dip ranges from 1580 Hz to 1930 Hz. Note that the undulation of transmission dip is mainly caused by the size limitation of the impedance tube system, which is attributed to the non-continuity of length for adjacent HST in the double layers HSTs. Theoretically, according to Eqs (5), considering the resonance frequency $\omega_c \propto \frac{1}{l}$, and assuming that the difference of the HST length tends to be infinitesimal, a low transmission ratio without undulations is acquired in a wide bandwidth.

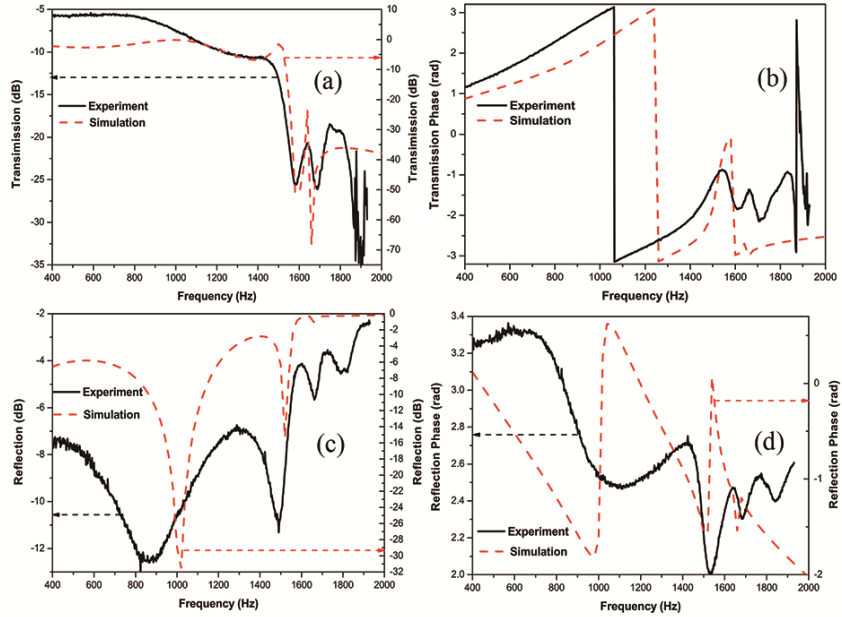

Figure 6: (a) Transmitted amplitude and (b) transmitted phase and (c) reflective ratio and (d) reflective phase of the double layers HSTs from the simulation and experiment.

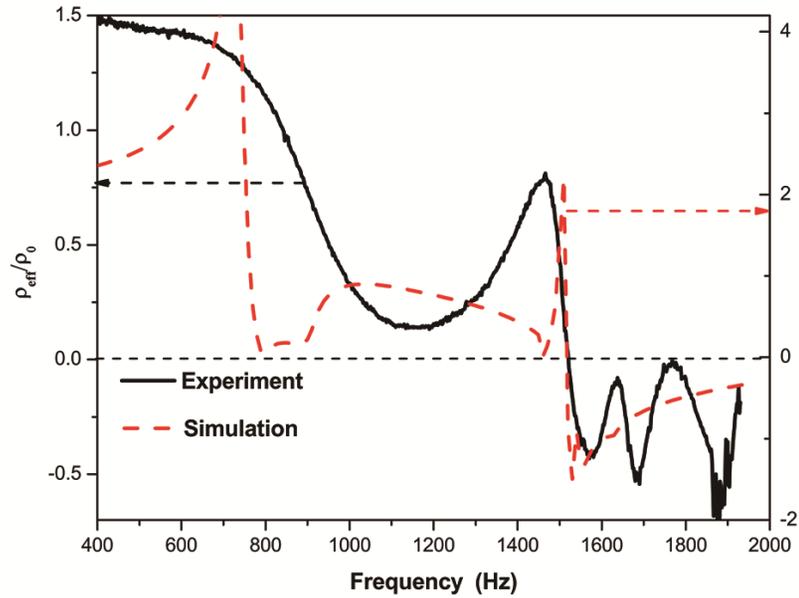

Figure 7: The effective mass density of the double layers HSTs.

Figure. 7 shows that the effective mass density of the 2D AM realized a negative value from 1520 Hz to 1930 Hz, which is the result of air resonance in the double layers HSTs.

## 3. The realization of double-negative parameters

The SHS structure [18, 19] is coupled with the HST. The SHS with a 12.5 mm radius, a 0.5 mm thickness, and a 3 mm split hole radius can achieve negative effective modulus with a resonance bandwidth that overlaps with the resonance frequency of the double layers HSTs. A 3D AM is fabricated by space ranking double layers SHS and double layers HSTs, as shown in Fig. 8. The SHS is periodically arrayed in the sponge matrix with a 30 mm lattice constant and a 40 mm thickness. The arrangements of the double layers HSTs are the same as those in Fig. 5(a). When the 3D AM is tested in the impedance tube system, the SHS layer faces the sound signal.

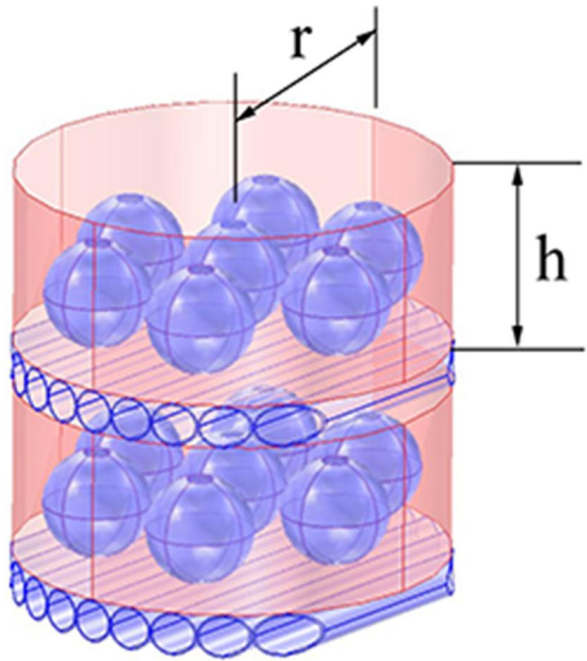

Figure 8: The schematic of the 3D AM.

Fig. 9 shows the transmission amplitude and phase curves of the 3D AM. The transmission ratio is shown to have a minimum value of -60.08 dB near 1520 Hz. As the frequency increases from 1520 Hz to 1750 Hz, the transmission ratio increases to a maximum value of -19.27 dB.

Simultaneously, the transmission phase has an inverse stage, in which the transmission ratio increases. Hence, the increase of the transmission ratio is ascribed to the matching of the resonance frequency of air in the HST and SHS.

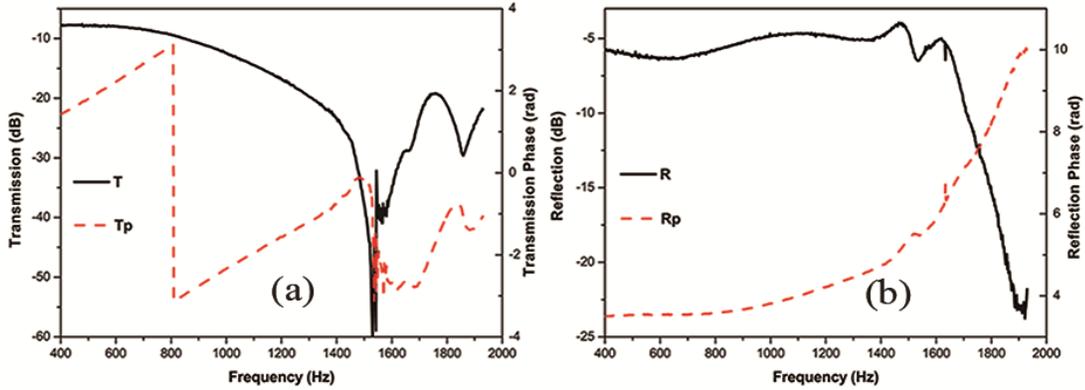

Figure 9: (a) Transmitted amplitude and phase and (b) reflective ratio and phase of the 3D AM.

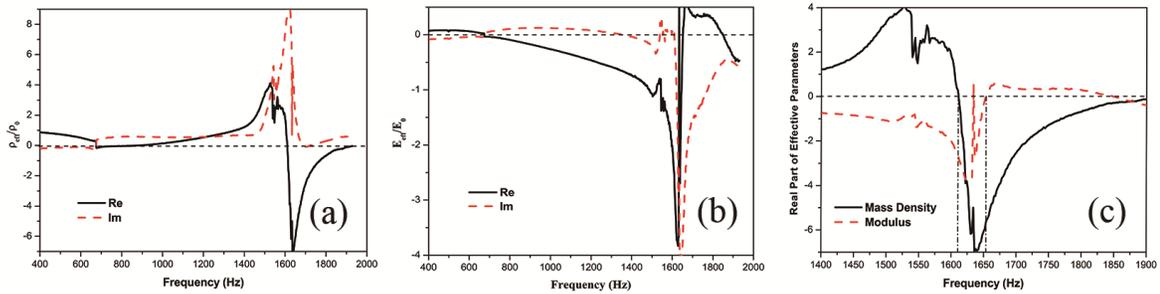

Figure 10: (a) The effective mass density and (b) effective modulus of the 3D AM. (c) The real part of double-negative parameters.

We obtain the effective mass density and effective modulus curves, as shown in Fig. 10. Obviously, in the bandwidth from 1612 Hz to 1654 Hz, the effective density and effective modulus achieve a double-negative.

## 4. Sub-wavelength imaging

The sub-wavelength imaging capabilities of the 3D double-negative AM is experimental measured. Three rectangles are perforated in an aluminum screen, as shown in Fig. 11(b), and placed in the front of the 3D AM. A speaker producing continuous sinusoidal waves is placed 8 cm before the aluminum screen. The operating frequency is chosen to be $f$=1630Hz, which corresponds to the wavelength $\lambda$=21.23 cm. The side length of the rectangle are $w$=1cm and $h$=4 cm with a centre-to-centre distance $s$=3 cm ($\lambda/7$). A microphone is placed 0.5 cm after the 3D AM to measure the 1D acoustic field distribution on the output side. Fig. 11(a) shows the normalized pressure amplitudes of the 3D AM with the three transmission peak, which demonstrate that the doube-negative AM has sub-wavelength absolution($\lambda/7$), in contrast to the measurement of sponge matrix with only a single peak.

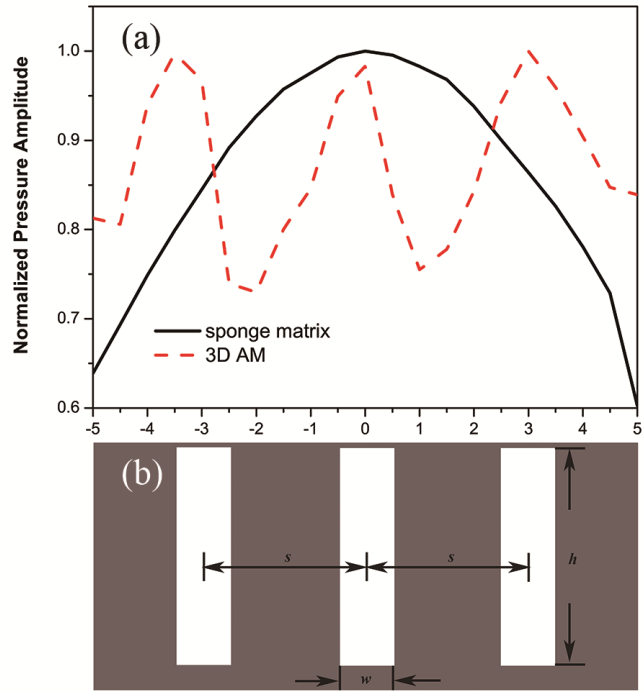

Figure 11: (a) Normalized pressure amplitudes in the image plane calculated at 1630Hz for sponge matrix and 3D AM. (b) The image source with w=1 cm, s=3 cm and h=4 cm.

## 5. Conclusion

Based on HST meta-atom model, a 2D AM with negative effective mass density was fabricated and the broadband negative effective mass density can be easily achieved by adjusting the length of HST. The transmission of HST AM shows that the transmission dip and inverse phase simultaneously appear near 1620 Hz. The FEM simulation results agree with the above experimental results and demonstrate the resonance mechanism of negative mass density. By coupling an HST and an SHS, we construct a meta-molecule model similar to the electromagnetic metamaterials that contains split ring resonators and metal cylinders. A 3D AM was fabricated, the experiment shows that the 3D AM obtains simultaneous the negative effective mass density and the negative effective modulus from 1612 Hz to 1654 Hz. The double-negative AM have been used to realize sub-wavelength imaging and its resolution reaches $\lambda/7$ at the resonance frequency 1630Hz.